\UseRawInputEncoding
\documentclass[pra,aps,twocolumn,superscriptaddress,showpacs]{revtex4}
\usepackage{graphicx}
\usepackage{amsmath}

\newcommand{\eq}{\begin{equation}}
\newcommand{\fine}{\end{equation}}

\def\be{\begin{equation}}
\def\ee{\end{equation}}
\def\bea{\begin{eqnarray}}
\def\eea{\end{eqnarray}}
\begin{document}

\title{   Bell Inequalities and Maximally Realistic Causal Quantum Mechanics}

\author{S. M. Roy}
\email{shasanka1@yahoo.co.in,smroy@hbcse.tifr.res.in} \affiliation{INSA Emeritus Scientist, Homi Bhabha Centre for Science Education, TIFR, V. N. Purav Marg, 
Mankhurd, Mumbai - 400 088.}

\begin{abstract}
The De Broglie-Bohm (DeBB)\cite{DeBB} Causal Quantum Mechanics played a crucial role in Bell's discovery \cite{Bell1964} that quantum mechanics 
violates EPR local reality \cite{EPR1935}, and also in Bell's search for an exact quantum mechanics. The experiments of Aspect et al
\cite{Aspect1981} confirm quantum correlations between plane polarizations of two photons and violation of Bell's inequalities 
by a factor $\sqrt 2 $. I prove that similar experiments with elliptic polarizers can also show  
 quantum  violations of Bell's inequality by the same factor. I summarize our construction of a maximally realistic causal quantum mechanics 
  in $n-$dimensional configuration space \cite{Roy-Singh1995}. Phase space Bell inequalities and 'Marginal Theorems' 
  \cite{Auberson2002} play a crucial role.
   
\end{abstract}


\maketitle

\section*{\large{1.  Introduction}}
Standard Quantum Field Theory incorporates the principle of 'no signalling faster than light' by means of the hypothesis that Field 
Observables at spacelike separation commute. Einstein, Podolsky and Rosen (EPR)\cite{EPR1935} enunciated a different principle  ,
 the ``Local Reality Principle'' or ``Einstein Locality``and used it to claim that description of the quantum mechanical state by the 
 wave function alone is incomplete. Bell 
  demonstrated \cite{BellHV} that the De Broglie-Bohm (DeBB)\cite{DeBB} Causal Quantum Mechanics ( which adds position as a hidden 
  variable with distribution identical to quantum position  probability density) explicitly violated local reality . He then made the 
 startling discovery that any theory which agrees with quantum correlations must violate Bell inequalities \cite{Bell1964}
 implied by Einstein locality. I summarize Bell's inequalities and
  experiments \cite{Aspect1981} on correlations of plane polarizations of photons which confirm quantum correlations and 
  violate Bell's inequalities by a factor $\sqrt 2 $. I suggest similar experiments with elliptic polarizers and show that 
  those quantum correlations also violate Bell's inequality by the same factor. The basic reason for quantum violation of Bell 
  inequalities is the context dependence of quantum probabilities. 'Local Reality' implies existence of joint probabilities 
  of some non-commuting observables which cannot be measured with one experimental arrangement. 
  
  Bell considered the DeBB causal theory as the most serious attempt towards an exact quantum mechanics because it avoids 
an ill-defined division of the world into quantum system and classical observer.However,
the DeBB theory fails to reproduce quantum probabilities of momentum \cite{Takabayasi}. Virendra Singh and I  
\cite{Roy-Singh1995} constructed a causal quantum mechanics in $n-$dimensional configuration space exactly reproducing the quantum probabilities of 
 $n+1$ complete commuting sets (CCS) of observables including position and momentum,as marginals of one positive 
phase space density . Auberson et al \cite{Auberson2002} derived phase space Bell inequalities and marginal theorems to show that the 
Roy-Singh causal quantum mechanics is maximally relistic: there exist quantum states for which no more than 
$n+1$ CCS of observables may be reproduced as marginals of one positive phase space density. There are exceptional states, viz. those 
with positive Wigner function \cite{Wigner1932} for which quantum probabilities of $2^n$ CCS agree with corresponding marginals 
of the Wigner function. Banaszek et al \cite{Banaszek} showed that even these quantum states can violate Bell's inequalities 
on correlations of certain operators such as displaced parity operators. This extends the context dependence seen in standard and maximally realistic 
causal quantum mechanics to observables non-diagonal in position or momentum.

Quantum states, unlike classical states, do not specify values of positions $q$ and momenta $p$ of particles. 
They only yield probabilities of finding values of $q$ (or $p$) if  $q$ (or $p$), were to be measured.
Joint probabilities of a complete commuting set (CCS) of observables are specified , but joint 
probabilities of any two non-commuting observables such as $q$ and $p$ are not. Further,
quantum states for a system consisting of two sub-systems cannot in general be written as products of individual  
wave functions for the two sub-systems but only as superpositions of such products. Such states are 
called entangled or non-separable \cite{Schrodinger}.

In standard quantum mechanics causality/determinism  does not hold. 
 Consider a quantum superposition $\alpha|z+\rangle +
\beta|z-\rangle$ for a spin-1/2 particle, where $|z+\rangle$ and
$|z-\rangle$ are eigenstates of $\sigma_z$ with eigenvalues +1 and -1
respectively.  When the same state is prepared repeatedly and passed
through a Stern-Gerlach apparatus to measure $\sigma_z$, 
\begin{equation}
 \alpha|z+\rangle +\beta|z-\rangle \>\rightarrow STERN-GERLACH\>\rightarrow \> \pm
\end{equation}
quantum mechanics cannot predict whether the result $+$ or the result $-$ will 
occur in a given trial; it merely says that a fraction
$|\alpha|^2$ of the particles ends up at the detector corresponding to
$\sigma_z = +1$, and a fraction $|\beta|^2$ goes to the detector
corresponding to $\sigma_z = -1$.  Thus different results (going to
one detector or the other) arise from exactly the same cause (the same
initial state).  Of course this lack of causality might be restored in
a theory in which the wave function is not a complete description of
the state of the system. 

Another striking example is that of a particle passing through a narrow hole on to a
hemispherical fluorescent screen. Although described by the same wave function spread nearly uniformly over the hemisphere,  
in repeated trials the particle arrives at different points on the screen.

Einstein in 1933 \cite{Einstein1933} expressed dissatisfaction with quantum theory being 
merely a set of rules about the statistics of measurement results :``I still believe in the 
possibility of giving a model of reality which shall represent events themselves and not merely 
the probability of their occurence'',  and more  specifically,\cite{Einstein1933} (p. 666) `` I am, in fact, rather firmly 
convinced that the essentially statistical character of contemporary quantum theory is solely to 
be ascribed to the fact that this (theory) operates with an incomplete description of physical 
systems ''.
  
Finally Einstein, Podolsky and Rosen \cite{EPR1935} posed the question ``Can Quantum Mechanical Description of Physical Reality 
be Considered Complete?''. Formulating and applying a 'local reality principle' on an entangled state of 
two spatially separated sub-systems $S_1$ and $S_2$ which have interacted in the past they concluded : ``While we have 
thus shown that the wave function does not provide a complete description of the physical reality, we left open the question 
of whether or not such a description exists. '' 

This open question  opened a treasure trove in front of John Bell \cite{BellHV} . He had just demolished 
all arguments against hidden variables in quantum mechanics \cite{Gleason}  as being unreasonable, and advocated  
the De Broglie-Bohm (DeBB) Causal Quantum Theory \cite{DeBB} as an explicit counterexample. DeBB added position as additional variable 
to the specification of the quantum state ,e.g. $\{\psi,x_1,x_2\}$ for a two particle system .
Bell considered this system in Stern-Gerlach type magnetic fields. He showed that in the case of EPR type entangled states the DeBB 
trajectory of particle 1 depends on the trajectory of 2 and hence on the analyzing fields acting on 2, however far these may be from particle 1 . 
Thus the DeBB theory, while agreeing with position predictions of usual quantum mechanics, provided an explicit causal mechanism of 
violating Einstein's Local Reality principle.

This led Bell to the next seminal question: must every hidden variable account of quantum mechanics have this 
extraordinary non-local character ? The answer, 'Yes', is given by Bell's theorem \cite{Bell1964} : every Local Hidden Variable (LHV) theory 
must violate quantum mechanics. Bell's theorem  is regarded by some as the most 
fundamental discovery of the 20th century \cite{Stapp}.Bell's theorem was later generalized by Wigner 
\cite{Wigner1970} who demonstrated a conflict of Einstein locality with quantum mechanics without explicit mention of hidden variables.

\section*{\large{2. Einstein's Principle of Local Reality and the EPR Paradox}}

{\bf Einstein Locality}. Suppose two systems $S_1$ and $S_2$ which have interacted in the past 
have now separated and are experimented upon by two observers in spatially separated regions. 
Observed properties of the two sub-systems can ofcourse be correlated due to past interactions.
Einstein insisted however on a local reality principle ,\cite{Einstein1933},p.85: 
``But on one supposition we should,in my opinion, absolutely hold fast: the 
real factual situation of the system $S_2$ is independent of what is done with the 
system $S_1$, which is spatially separated from the former.''  

The meaning of 'real factual situation' in Einstein's formulation is 
clarified by the definition of 'physical reality' in the landmark paper of 
Einstein, Podolsky and Rosen \cite{EPR1935}: `` If without in any way disturbing a system, we can 
predict with certainty (i.e. with probability equal to unity) the value of a physical quantity,
then there exists an element of physical reality corresponding to this physical quantity.'' 
In this sense ,in quantum mechanics an observable has physical reality only in it's eigen states. 

Einstein, Podolsky and Rosen applied the principle of local reality to 
argue that Quantum Mechanics is incomplete. Consider a two particle system. 
In quantum mechanics, the position observable $\hat q_i$ and the momentum observable 
$\hat p_i$ cannot be simultaneously specified sharply; but the observables $\hat q_1 - \hat q_2$ and 
$\hat p_1 + \hat p_2$ are commuting observables and there is a quantum state
\be
|\hat q_1 - \hat q_2 =q_0 \rangle |\hat p_1 + \hat p_2 =p_0 \rangle  \label{EPR state}
\ee

in which they  
are specified arbitrarily sharply, and have values $q_0$ and $p_0$. Ignore momentarily 
the difficulty that such states are not normalizable, a difficulty removed later 
by Bohm and Aharonov by considering spin observables. Suppose observers A and B are 
spacelike separated. In such a state, if B chooses to measure $q_2$ she predicts $q_1$ 
with certainty (without disturbing that particle since it is spatially separated), and 
hence $q_1$ must have physical reality; equally, if she chooses to measure $p_2$ she 
predicts $p_1$ to have reality. By the principle of local reality, reality for particle 1 
must be independent of choices made by observer B. Hence, both $q_1$ and $p_1$ must have 
physical reality for particle 1. No quantum state allows simultaneously sharp $q_1$ and $p_1$.
Thus, EPR conclude that quantum theory must be incomplete. 

EPR envisaged that an extension (or completion ) of quantum mechanics agreeing with local reality 
would be possible. Bell's theorem  shattered this hope .

\section*{\large{3. EPR Experiment, Local Hidden Variables and Bell's Theorem}}

\begin{center}
 \includegraphics[width=.55\columnwidth]{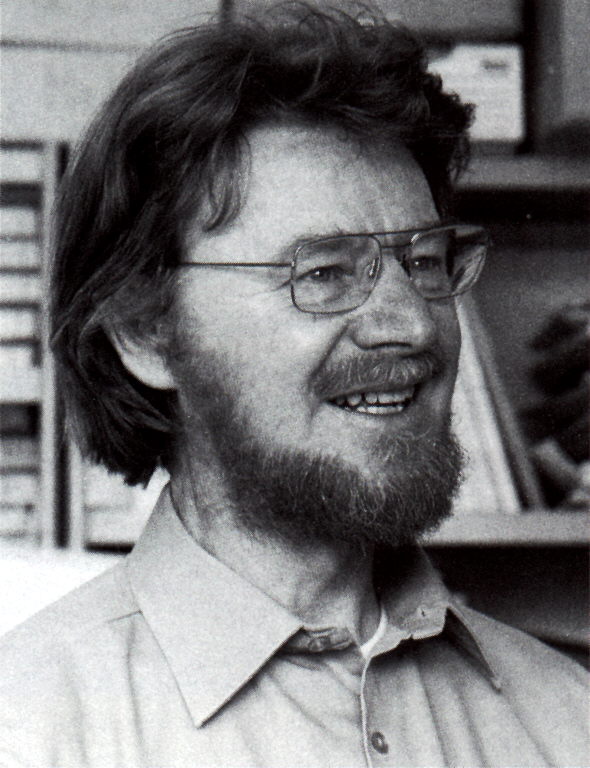}\\
John S. Bell
\end{center}

\begin{figure}[ht]
 \begin{center}
 \includegraphics[width=.75\columnwidth]{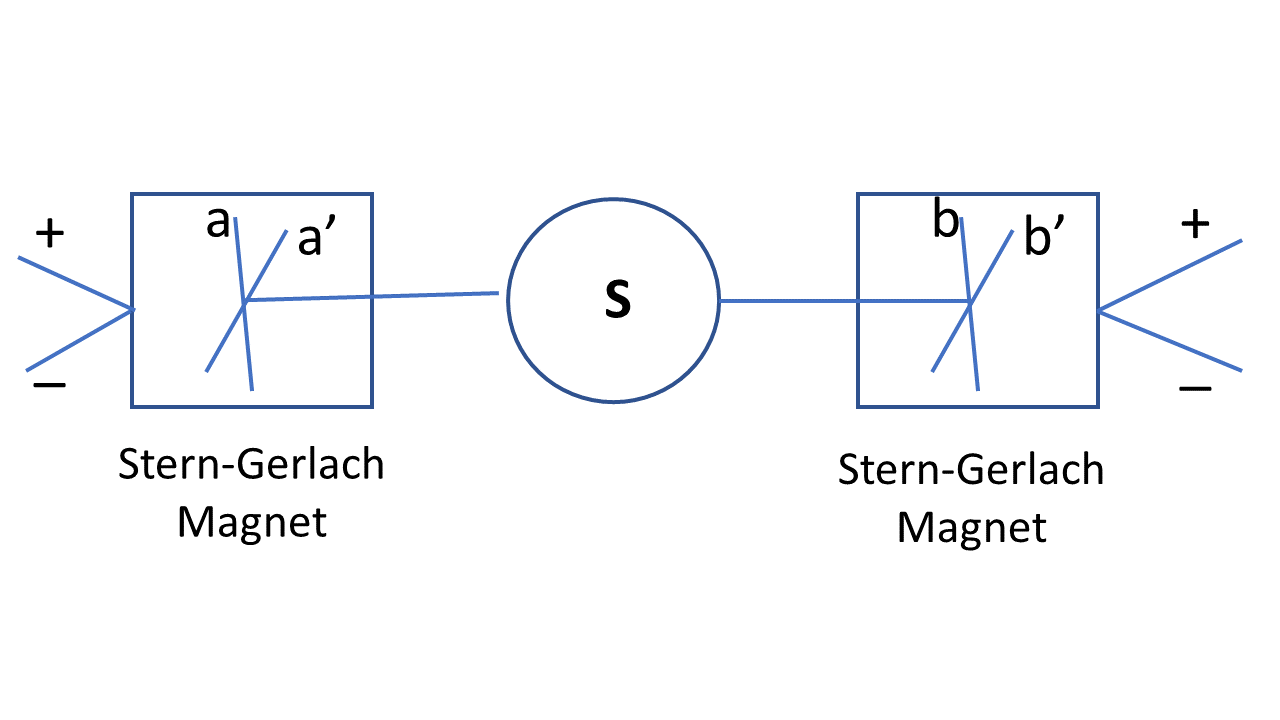}\\
\end{center}
\caption{EPR-Bohm-Aharonov Experiment}
\end{figure}

\begin{figure}[ht]
\begin{center}
 \includegraphics[width=.55\columnwidth]{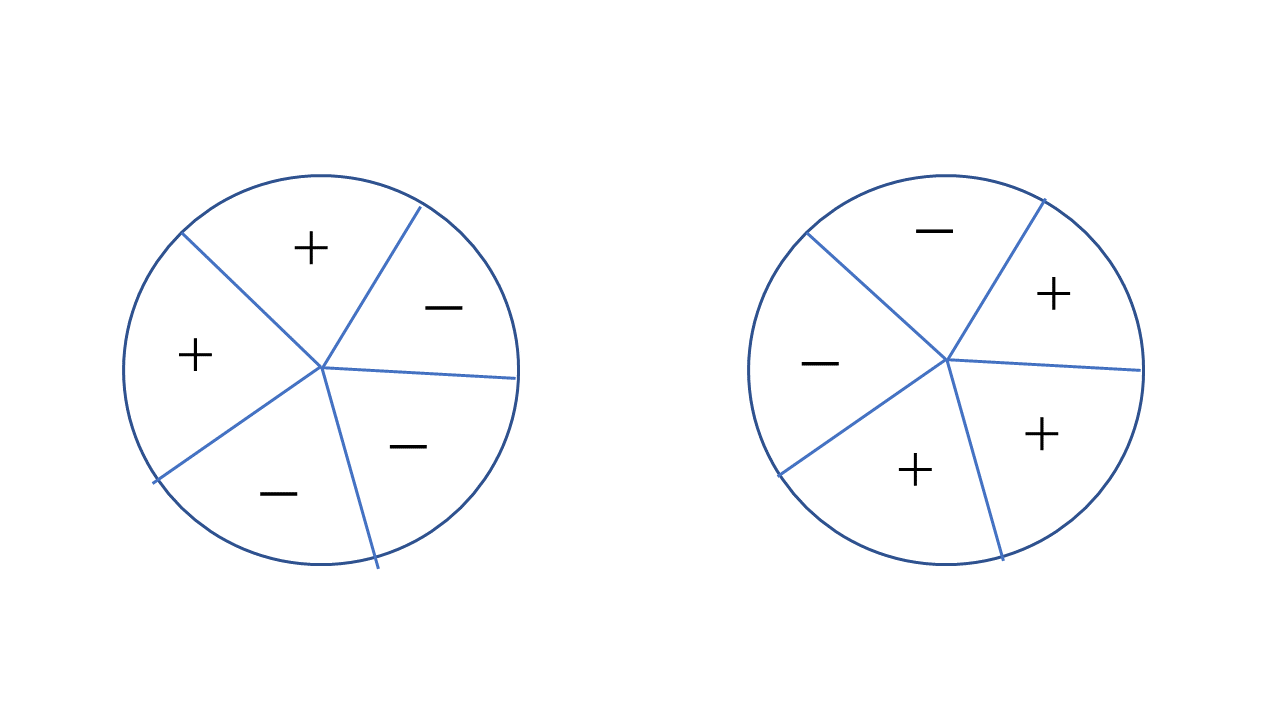}\\
 \end{center}
\caption{Classical Explanation of EPR-Anticorrelations}
\end{figure}

Bell formulated the EPR local reality idea for two particles emitted in opposite directions, using correlations $ \langle AB\rangle $ 
between measured values $A=\pm 1$ for one particle and $B=\pm 1$ for  the other. Although the idea is primarily meant for 
relativistic particles such as photons, the first theoretical application was to the non-relativistic Bohm-Aharonov \cite{Bohm-Aharonov} 
example of two spin-half particles or two qubits (quantum bits) prepared in a singlet state
\begin{equation}
 |\Psi \rangle = |\uparrow \downarrow -\downarrow \uparrow \rangle / \sqrt{2} 
   =|\nwarrow \searrow -\searrow \nwarrow \rangle / \sqrt{2}
\end{equation}
at a source $S$ and then flying apart. Two  observers, each 
equipped independently, (e.g. with rotatable Stern-Gerlach magnets)  measure arbitrary
components of the particle spin  chosen during the flight of the particles such that the measurements are spacelike separated.
Since the singlet state is rotationally symmetric, the spin components measured along any direction by the two observers must be 
opposite. Observer 1 can predict with certainty the result of measurement of 
any component of $\vec{\sigma}^{(2)}$.$\vec{a}$ by observer 2 by previously  
measuring the same component $\vec{\sigma}^{(1)}$.$\vec{a}$ for particle 1.Since Einstein locality implies 
that the choice of magnet orientation made by the remote observer 1 does not affect the result 
obtained by observer 2, the result of any such observation must be predetermined.The 
initial quantum wave function does not specify the result of an individual measurement;therefore 
the predetermination requires a more complete specification of the state , say by adding 
additional variables ('hidden variables') $\lambda $.

Hidden variables achieving perfect anti-correlation 
(in each individual shot) between $+$ and $-$ results along the same direction poses no problem for 
classical visualization. E.g. two  
discs could be shot off along opposite directions with initial markings $+$ and $-$ on the two discs being anticorrelated for every direction.
We may even allow probabilistic rather than deterministic hidden variables. Bell showed however, that the imperfect correlations predicted by 
quantum mechanics conflict with local reality.
Suppose the hidden variables $\lambda $ , with probability distribution $\rho (\lambda ) $ ,determine 
the probability $p_1^r(\lambda ,\vec{a}) $ of observing the value $r=\pm 1$ of 
$A(a)$ corresponding to the quantum observable $\vec{\sigma}^{(1)}$.$\vec{a}$ and the probability $p_2^s(\lambda ,\vec{b}) $ of observing 
the value $s=\pm 1$ of $B(b)$ corresponding to the quantum  observable $\vec{\sigma}^{(2)}$.$\vec{b}$. {\bf Einstein locality} allows only 
Local Hidden Variables (LHV) such that $p_1^r(\lambda ,\vec{a}) $ and $p_2^s(\lambda ,\vec{b}) $ are independent of the orientations of the 
remote measuring apparatus.Then the probability $p^{r,s}(a,b) $ of observing $A(a)=r $ and $B(b)=s $ is,
\begin{equation}
 LHV: p^{r,s}(a,b)= \int d \lambda \rho (\lambda ) p_1^r(\lambda ,\vec{a})p_2^s(\lambda ,\vec{b}) ,
\end{equation}
and the correlation $P(a,b)=<A(a) B(b)>$ is predicted to be,
\begin{eqnarray}
&&P(a,b)=\sum _{r=\pm 1,s=\pm 1 }rs \> p^{r,s}(a,b)\nonumber\\
&&=\int d \lambda \rho (\lambda ) \bar{A} (\lambda, a)\bar{B} (\lambda, b), \>where,\>\nonumber \\
&&\bar{A} (\lambda, a)= p_1^+(\lambda ,\vec{a})- p_1^-(\lambda ,\vec{a}) ,\>|\bar{A} (\lambda, a)| \leq 1,\nonumber\\
&&\bar{B} (\lambda, b)=p_2^+(\lambda ,\vec{b})- p_2^-(\lambda ,\vec{b}),\>  |\bar{B} (\lambda, b)|\leq 1.\nonumber
\end{eqnarray}
  These LHV correlations were shown to conflict with the quantum values $P_{QM}(a,b)=<\vec \sigma_1
\cdot \vec a \ \vec \sigma_2 \cdot \vec b >$ ,when we consider four possible measurements, 
with two orientations $a,a'$ on one side and two orientations $b,b'$ on the other side .

{\bf Wigner's proof of Bell's Theorem}. Without any restriction to non-relativistic contexts, Bell's hypothesis actually allows the 
construction of a joint probability 
\begin{eqnarray}
&& p^{r,r',s,s'}(a,a',b,b')=\int d \lambda \rho (\lambda ) p_1^r(\lambda ,\vec{a})p_2^s(\lambda ,\vec{b})\nonumber\\
&& \times p_1^{r'}(\lambda ,\vec{a}')p_2^{s'}(\lambda ,\vec{b}')
\end{eqnarray}
for $A(a),A(a'),B(b), B(b')$ to have the values $r,r',s,s'$ respectively, such that the result of any of the four 
feasible experiments can be obtained as a ``marginal''. E.g.
\begin{equation}
  p^{r,s}(a,b)= \sum _{r'=\pm 1,s'=\pm 1 }\> p^{r,r',s,s'}(a,a',b,b').
\end{equation}
The experimental correlation $\langle A(a) B(b)\rangle $ is then,
\begin{equation}
P(a,b)=\sum _{r=\pm 1,s=\pm 1 }rs \> p^{r,s}(a,b)
\end{equation}
with similar expressions for $P(a',b),P(a,b'),P(a',b') $. Hence, using the positivity of probabilities,
\begin{eqnarray}
&|P(a,b)-P(a,b')| +|P(a',b)+P(a',b')|\nonumber\\
&\leq \sum _{r,s,r',s'}\big (|r(s-s')|+|r'(s+s')| \big ) p^{r,r',s,s'}(a,a',b,b')\nonumber \\
&=2 \label{Bell-CHSH}
\end{eqnarray}
which is the Bell-CHSH \cite{Bell1964} local reality inequality {\bf without any restriction to non-relativistic contexts}. 
Here we used $|r(s-s')|+|r'(s+s')|=2 $ and the requirement that the joint probability summed over all values of $r,s,r',s'$ must be unity.

In contrast, Quantum Mechanics gives,
\begin{equation}
 P_{QM }(a,b)= -\vec{a}.\vec{b}.
\end{equation}
The choice of coplanar vectors such that $\vec{a}.\vec{b}=-\vec{a}.\vec{b'}=\vec{a'}.\vec{b}= \vec{a'}.\vec{b'}
=1/\sqrt{2}$ yields the value $2\sqrt{2} $ for the left-hand side of the Bell-CHSH inequality in gross 
violation of local reality! .It has been proved \cite{Cirelson} that this is the maximum violation possible for two qubit states.
It has also been proved that every two qubit entangled pure state must violate a Bell-CHSH inequality \cite{Gisin}.

{\bf Multiple settings, multiparticles,Local Reality versus Separability }
The Bell-CHSH inequalities use two settings at each site and are not sufficient to guaranty local reality: independent 
local reality inequalities with $M$ settings at one site and $N$ settings at the other have been obtained \cite{Roy-Singh1978}. For $N$ qubits 
but two settings at each detector, local reality Bell inequalities may be violated 
 by a factor $2^{(N-1)/2}$ \cite{mermin1990},\cite{roy-singh1991} . For quantum computation, the more relevant property is 'entanglement' or its opposite, 
viz. 'separability'. Quantum separability has been shown 
to imply inequalities exponentially stronger than local reality Bell inequalities for large $N$ \cite{Roy2005}. 
They are violated by  a factor $2^{N-1}$ by some entangled states; these states are 
natural candidates to exploit in seeking improvements over classical computation.

\begin{figure}[ht]
 \begin{center}
 \includegraphics[width=1.0\columnwidth]{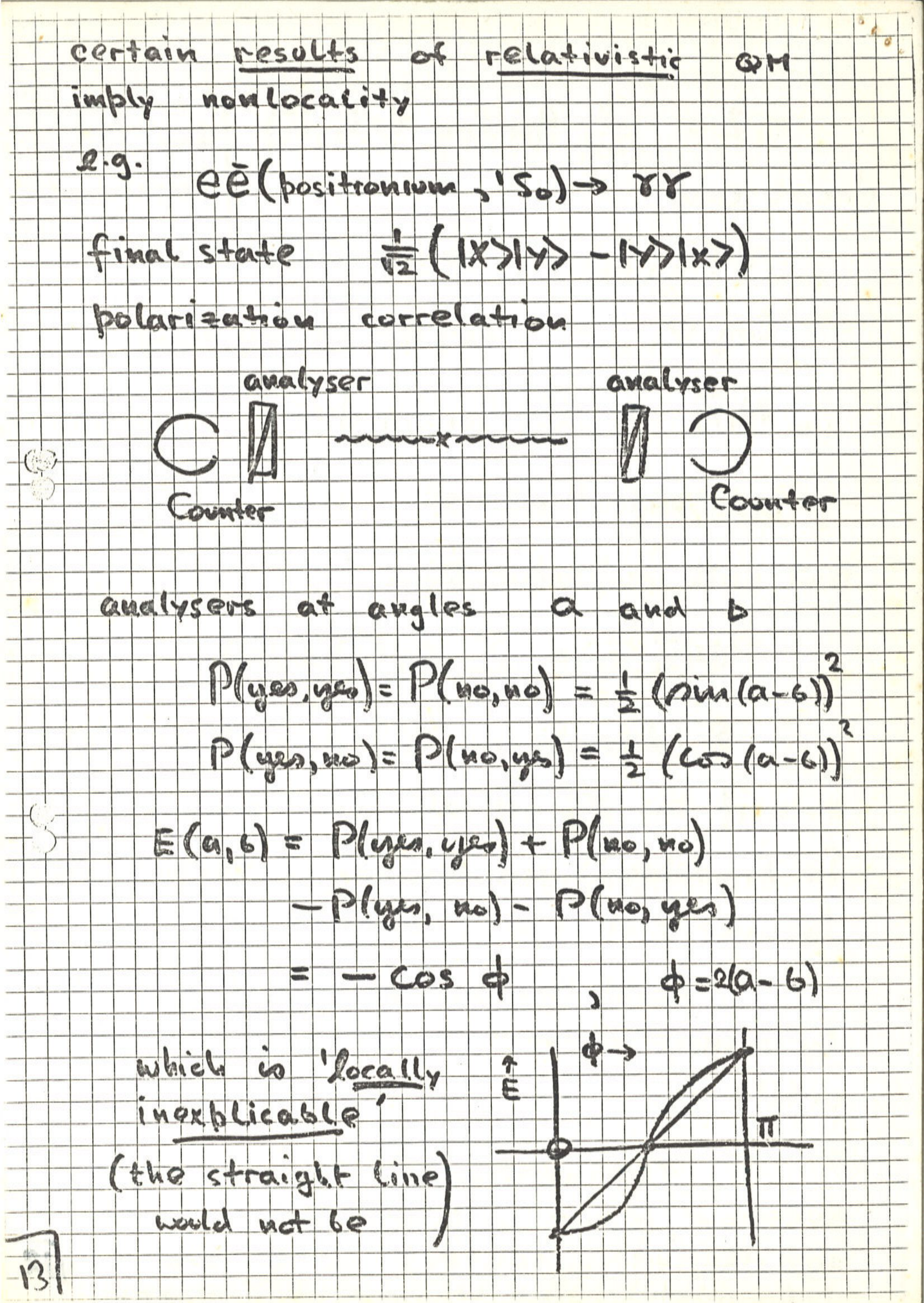}
\end{center}
\caption{Transparency 13 of Bell's TIFR lecture (1982)}
\end{figure}

\section*{\large{4. EPR-Bell Experiments on Two Photons with Linear and Elliptic Polarizers }}
We come now to relativistic tests.
 Measurements of plane polarizations on two photon states suggested by Bell\cite{Bell1975} and carried out by 
 Alain Aspect and others \cite{Aspect1981}  verified the  quantum predictions, and disproved  the EPR local reality hypothesis.
I summarize their results and also propose analogous  EPR-Bell experiments using elliptic polarizers/analyzers. 
I show that quantum correlations in the proposed experiments with elliptic polarization detectors on 
both sides violate Bell-CHSH inequalities by a factor $\sqrt 2$ ,but experiments detecting plane polarization on one side 
and elliptic polarization on the other do not violate local reality inequalities. 

The decay of the $^1S_0$ state of Positronium, or of $\pi^0 $ yields the two photon state of zero angular momentum
\begin{eqnarray}
 |\Psi \rangle _- = (|x \rangle |y \rangle  -|y \rangle |x \rangle )/ \sqrt{2} \\
   = i(|+\rangle |- \rangle  -|- \rangle |+ \rangle ) / \sqrt{2} ,\nonumber\\
   |\pm\rangle \equiv (|x\rangle \pm i|y\rangle )/\sqrt{2},
\end{eqnarray}
where $ |x \rangle ,|y \rangle$ denote photons polarized in $x$ , $y$ directions and $|\pm \rangle $ are circularly polarized photons.
Experiments on atomic cascade photons \cite{Aspect1981} have used the state,
\bea
 |\Psi \rangle _+= (|x \rangle |x \rangle  +|y \rangle |y\rangle )/ \sqrt{2} \nonumber\\
 = (|+\rangle |- \rangle  +|- \rangle |+ \rangle ) / \sqrt{2} . 
\eea
For both $ |\Psi \rangle _\pm $, as the two photons travel in opposite directions, they have the same circular polarization 
and the total spin angular momentum is zero. The states can be rexpressed in the basis of states $|\theta \rangle$ plane polarized 
in arbitrary directions, and corresponding elliptically polarized states $|\theta \rangle _E$,
\begin{eqnarray}
 |\theta \rangle \equiv cos \theta |x\rangle + sin \theta |y \rangle ;\nonumber\\
 |\theta \rangle_E \equiv cos \theta |x\rangle + i sin \theta |y \rangle .
\end{eqnarray}
Then,
\begin{eqnarray}
 |\Psi \rangle _- = (|\theta \rangle |\theta +\pi/2 \rangle  -|\theta +\pi/2\rangle |\theta\rangle )/ \sqrt{2} \nonumber\\
 =-i (|\theta \rangle_E |\theta +\pi/2 \rangle_E  -|\theta +\pi/2\rangle_E |\theta\rangle_E )/ \sqrt{2} ;
\end{eqnarray}
and
\begin{eqnarray}
 |\Psi \rangle _+ = (|\theta \rangle |\theta \rangle  +|\theta +\pi/2\rangle |\theta +\pi/2\rangle )/ \sqrt{2} \nonumber\\
 =-(|\theta \rangle_E |-\theta  \rangle_E + |\theta +\pi/2\rangle_E |-\theta -\pi/2 \rangle_E / \sqrt{2} ;
\end{eqnarray}

For both $|\Psi \rangle _\pm$  a plane polarization measurement on one photon forces the other into a plane polarized state,
 and an elliptic polarization measurement on one forces the other photon into an elliptic  polarization state. EPR locality says that the choice of 
measurement on one photon cannot affect the real situation of the distant second photon ; this would have the paradoxical implication 
that the second photon is both plane polarized and elliptically polarized !

Consider now a generalized Bell-type experiment (see Fig. 3 )in which the initial state can be $ |\Psi \rangle _\pm $, and 
analyser settings $a,a'$ and $b,b'$ can be chosen to transmit plane or elliptically polarized 
photons $|\theta \rangle$ or $|\theta \rangle _E$ ,with any desired values of $\theta$.
Suppose $A_{L,E} (a)=\pm 1$ and $B_{L,E}=\pm 1$ correspond to transmission or non-transmission of the photon by the left-analyzer 
and right-analyzer respectively, the subscripts $L,E$ corresponding to Linear and Elliptic polarization. 
If the left analyzer is set to transmit $|\theta \rangle _E=|a \rangle _E $ photons, and the right analyzer is set to transmit 
$|\theta \rangle _E=|b \rangle _E $ photons, noting that $_E\langle a |a+\pi/2 \rangle _E =0 $,we have

\begin{eqnarray}
 & &(P_{E,E}(a,b))_{\pm } =_\pm\langle \Psi |A_E(a) B_E(b)|\Psi \rangle _\pm \nonumber \\
 &=&_\pm\langle \Psi |\big(|a\rangle \langle a| -|a+\pi/2\rangle \langle a +\pi/2|\big)_E \nonumber \\
 &\times & \big(|b\rangle \langle b| -|b+\pi/2\rangle \langle b +\pi/2|\big)_E |\Psi \rangle _\pm\nonumber \\
 &=& \pm cos (2(a\pm b)) .
\end{eqnarray}
and 
\begin{eqnarray}
 & &(P_{L,L}(a,b))_{\pm } =_\pm\langle \Psi |A_L(a) B_L(b)|\Psi \rangle _\pm \nonumber \\
 &=&_\pm\langle \Psi |\big(|a\rangle \langle a| -|a+\pi/2\rangle \langle a +\pi/2|\big) \nonumber \\
 &\times & \big(|b\rangle \langle b| -|b+\pi/2\rangle \langle b +\pi/2|\big) |\Psi \rangle _\pm\nonumber \\
 &=& \pm cos (2(a - b)) .
\end{eqnarray}
Similarly ,
\begin{eqnarray}
 & &(P_{L,E}(a,b))_{\pm } =_\pm\langle \Psi |A_L(a) B_E(b)|\Psi \rangle _\pm =\pm cos 2a cos 2b\nonumber \\
 & &(P_{E,L}(a,b))_{\pm } =_\pm\langle \Psi |A_E(a) B_L(b)|\Psi \rangle _\pm =\pm cos 2a cos 2b \nonumber
 \end{eqnarray}
This gives,
\begin{eqnarray}
 |P_{E,E}(a,b)-P_{E,E}(a,b')|_\pm + \nonumber \\
 |P_{E,E}(a',b)+P_{E,E}(a',b')|_\pm=2\sqrt 2 ,
\end{eqnarray}
if
\begin{equation}
 \{2a,2b,2a',2b'\}=\{0,\pi/4,\pi/2,3\pi/4 \},
\end{equation}
thus violating the Bell-CHSH \cite{Bell1964} local reality inequality 
 by a factor $\sqrt 2$, just as for linear polarization correlations $(P_{L,L}(a,b))_{\pm }$ . No violation is predicted for $(P_{L,E}(a,b))_{\pm }$ and 
$(P_{E,L}(a,b))_{\pm }$ with linear polarizers on one side and elliptic polarizers on the other.
We expect that experiments with elliptical polarizers 
will give new confirmation of quantum mechanics and strong violation of local reality.

\section*{\large{5. Causal quantum theory a step towards an exact quantum theory? }}

 John Bell's seminar at TIFR in 1982, 'What in the world is quantum mechanics about exactly ?', detailed exposition 
 `Towards An Exact Quantum Mechanics'  \cite{BellSchwingerFest1989}, and lectures `against measurement' 
\cite{Erice1989} exemplify his relentless search for an exact quantum theory.
Bell considered a quantum mechanics which gives nothing except the statistics of 
measurement results to be fundamentally ambiguous because it requires an 
undefinable boundary between the quantum object and the classical measuring apparatus.
`` Nobody knows what quantum mechanics says {\bf exactly} about any situation. For nobody knows 
where the boundary really is, between wavy quantum system and the world of particular events.
{\bf This is the problem of quantum mechanics}''  \cite{BellSchwingerFest1989}.
Bell repeatedly emphasized that 'the de Broglie-Bohm picture disposes of the necessity to divide the world 
somehow into system and apparatus' \cite{Bell1984}. He constructed a relativistic version 'beables for quantum field theory' 
which was a 'quantum field theory without observers,or observables,or measurements,or systems, or apparatus, or wave function collapse,
or anything like that'\cite{Bell1986}. 

The reason the DeBB causal theory does not need measurements is that the state is specified by 
the wave function plus additional or hidden variables, the positions of all the particles in the world,$\{|\psi \rangle (t),\vec x (t) \}$.
For non-relativistic particles ,it prescribes the  phase space density,
\begin{equation}
\rho_{dBB} = |\psi(\vec x,t)|^2 \delta(\vec p - \vec p_{dBB} (\vec
x,t)),
\label{thirteen}
\end{equation}

where $\vec p_{dBB} =  \vec\nabla S(\vec x,t)$ and $\psi = R \exp(iS/\hbar)$, with $R$ and $S$ real.  For any ensemble 
of configuration space points whose density is equal to the quantum position probability density at $t=0$, the continuity equation following from the 
Schr\"odinger equation ensures that the equality holds for all time.  However Takabayasi 
\cite{Takabayasi} pointed out that the quantum momentum probability density is not reproduced,

\[
\int \rho_{dBB} d\vec x \neq |\langle \vec p|\psi(t)\rangle|^2.
\]

{\bf Maximally Realistic Causal Quantum Mechanics}. 
This unsatisfactory feature of the DeBB theory \cite{Holland}, the breaking of the fundamental symmetry between  position and momentum 
can be removed.
Standard Quantum mechanics does not specify joint probabilities of noncommuting observables. For any complete commuting set (CCS) of observables A, the 
quantum state $|\psi \rangle$ specifies the probability of observing the eigenvalues $\alpha$ as 
$| \langle \alpha|\psi \rangle |^ 2$ , if A were to be measured. If B is another CCS with eigenvalues $\beta$, but 
$[A, B]$ is non-zero, the analogous probabilities $| \langle \beta|\psi \rangle |^ 2$   refer 
to a different context or experimental situation where  B is measured . Each context corresponds to the experimental arrangement to measure 
one CCS of observables, because non-commuting A and B cannot be accurately measured simultaneously,

Can we do better in  causal quantum mechanics? Can we remove the asymmetry between position and momentum?
In $n$ dimensional configuration space  Singh and I \cite{Roy-Singh1995} were able to construct a
new causal quantum mechanics in which the quantum probability
densities of $n+1$ CCS including $\vec x$ and $\vec p$ are simultaneously reproduced as marginals of
one positive definite phase space density.  The individual
trajectories are given by a c-number causal Hamiltonian and hence the phase 
space density is constant along the trajectory (Liouville property).We also conjectured that 
there exist states for which it is impossible to reproduce probabilities of more than $n+1$ CCS as
marginals of a positive definite phase space density. This conjecture was proved by  
Auberson et al \cite{Auberson2002} using phase space Bell inequalities .
In this sense the new mechanics is  ``maximally realistic''.
For $n=1$, Roy and Singh \cite{Roy-Singh1995} constructed two
simple phase space densities  which have this property,
\bea
& &\rho (x, p, t) = |\psi (x, t)|^2 \delta (p - \hat p (x, t))\nonumber\\
& &=| \tilde\psi (p, t) |^2 \ \delta (x - \hat x (p,t)),\nonumber\\
& &\>if\>|\psi (x, t)|^2 = \left| \frac{\partial \hat p (x, t)}{\partial x}\right| |\tilde\psi (p, t)|^2 .
\eea
Equivalently,
\bea
\rho(x,p,t) = |\langle x|\psi(t)\rangle|^2 |\langle p|\psi(t)\rangle|^2 \nonumber\\
\times \delta\left(\int^p_{-\infty} dp'|\langle p'|\psi(t)\rangle|^2 - 
 \int^{\epsilon x}_{-\infty} dx' |\langle \epsilon x'|\psi(t)\rangle|^2\right),
\label{fifteen}
\eea
where $\epsilon=\pm 1$ correspond respectively to  $\hat p (x, t) $ being non-decreasing and non-increasing functions.
This phase space density corresponds to
evolution of position and momentum according to a $c$-number causal
Hamiltonian of the form
\begin{equation}
H_c (x,p,t) = \frac{1}{2m} (p - A (x,t))^2 + V (x,t) ,
\end{equation}
with both $A(x,t)$ and $V(x,t)$ having parts which depend on the wave
function. Thus there are two quantum potentials instead of just one
in the DeBB theory. 

  However, to reproduce quantum probabilities for another pair of canonically conjugate observables 
which are linear combinations of position and momentum ,an analogous but  different phase space density is needed. 
This is an example of context dependence surviving in the new causal mechanics. 

In higher dimensions there is a pleasant surprise: $2^n$ phase space densities, each reproducing quantum probabilities of $n+1$ CCS 
can be constructed. E.g. for $n=2$, we can build four phase space densities ( $\epsilon_1=\pm 1, \epsilon_2=\pm 1$ )
 for each of which the quantum probability
densities of three different complete commuting sets (CCS) of
observables, e.g. $(X_1, X_2), (P_1, X_2), (P_1, P_2)$ is
simultaneously realized as marginals. Explicitly, the positive definite phase space
density
\bea
\rho (\vec x, \vec p, t) &=& |\psi (x_1, x_2, t)|^2 |\psi (p_1, x_2,
t)|^2 |\psi (p_1, p_2, t)|^2 \nonumber\\
& & \times \delta (A_1) \delta (A_2)
\label{density} 
\eea
where 
\begin{eqnarray} 
A_1 &\equiv& \int^{p_1}_{-\infty} |\psi (p'_1, x_2, t)|^2 dp'_1 -
\int^{\epsilon_1 x_1}_{-\infty} |\psi (\epsilon_1 x'_1, x_2, t)|^2 dx'_1 , \nonumber\\
A_2 &\equiv& \int^{p_2}_{-\infty} |\psi (p_1, p'_2, t)|^2 dp'_2 -
\int^{\epsilon_2 x_2}_{-\infty} |\psi (p_1,\epsilon_2 x'_2, t)|^2 dx'_2 \nonumber
\end{eqnarray}
reproduces as marginals the correct quantum probability densities
$|\psi (x_1, x_2, t)|^2 , |\psi (p_1, x_2, t)|^2$ and $|\psi (p_1,  
p_2, t)|^2$. (Note that $\psi (p_1, x_2, t) $ and $\psi (p_1, p_2, t)$ denote appropriate Fourier transforms 
of $\psi (x_1, x_2, t)$). 

{\bf Context dependence }.Replacing $\psi (p_1, x_2,t) $ by $\psi (x_1, p_2,t) $ in Eqn. \ref{density} 
we obtain a {\bf different} phase space density which reproduces quantum probability densities of $(X_1, X_2), (X_1, P_2), (P_1, P_2)$. 
The difference illustrates that context dependence, although less severe than in standard quantum mechanics, persists 
in causal quantum mechanics. 

{\bf Experimental tests}. This is largely unexplored. A preliminary idea is that the different experimental arrangements corresponding 
to realization of the different phase space densities may involve theories of approximate joint measurements of non-commuting 
observables \cite{Arthurs-Kelly}. E.g. for $n=1$ ,suppose the 
system with position observable $\hat q$ (with conjugate $\hat p$) interacts with an apparatus which has two commuting observables 
$\hat x_1,\hat x_2$ (with conjugate operators $\hat p_1,\hat p_2$ 
with the Arthurs-Kelly interaction $H_{A-K}=K(\hat q \hat p_1+\hat p \hat p_2)$ during time $T$ such that $KT=1$.Then 
approximate values of system position $q$ and momentum $p$ can be extracted from accurate observation 
of $x_1, x_2$ ,obeying $\langle q-x_1  \rangle =0,\langle p-x_2  \rangle =0 $ and
\be
(\Delta x_1 )^2 = (\Delta q )^2 + b^2, \>(\Delta x_2 )^2 = (\Delta p )^2 + \frac {1}{4 b^2},
\ee
where $b/\sqrt 2$ is the initial uncertainty of apparatus variable $x_1$ prepared in a Gaussian state.
Hence, for good approximation of both $q$ and $p$ distributions, we also need,
\begin{equation}
 \Delta q \Delta p \gg 1 ,\>(\Delta p)^{-1}\ll b\ll \Delta q\>.
\end{equation}
  A critical test of the $R-S$ distribution with $\epsilon=1$ against the $A-K$ distribution will be to 
compare the positions of the momentum peaks $p_{R-S} (q)$ and $p_{A-K}(q)$. This has been performed by 
Arunabha Roy \cite{Arunabha}  for the spreading free particle Gaussian wave function of the single slit case (see Fig. 4).
\begin{figure}[ht]
\begin{center} 
 \includegraphics[width=1.0\columnwidth]{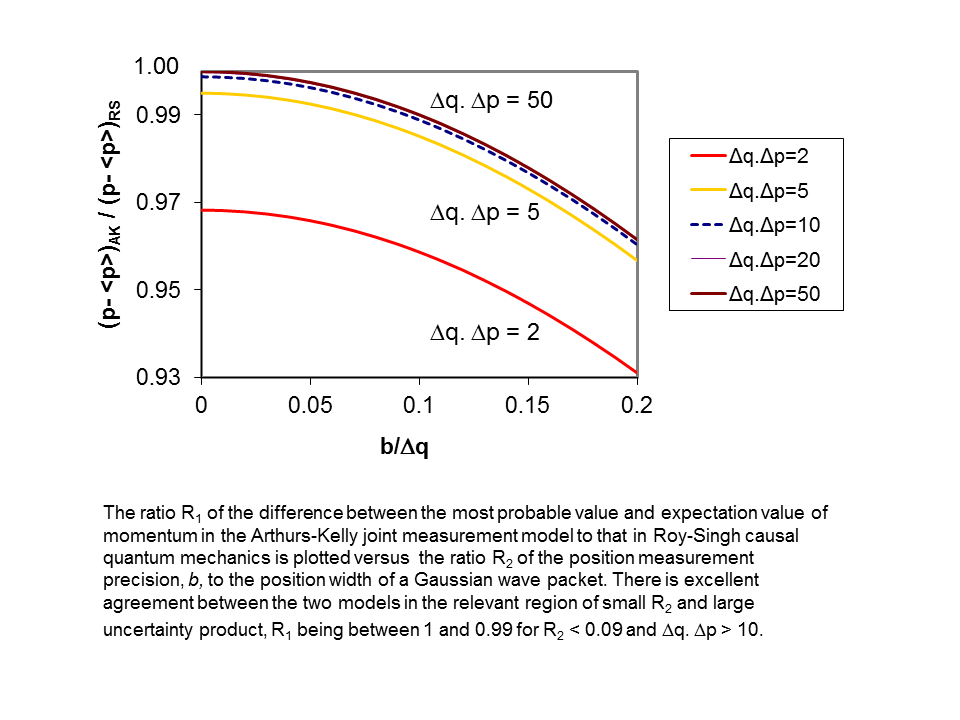}
\end{center}
\caption{Comparison of Arthurs-Kelly Joint measurement with Roy-Singh causal quantum-mechanics. Courtesy: A. S. Roy.}
\end{figure}
 Such tests for $n \ge 2$ might extend our understanding of context dependence.

{\bf Relativistic extensions} Extensions of the DeBB theory to Bose \cite{DeBB} and Fermi fields \cite{Bell1986} exist.
Such extensions for the maximally realistic causal theory are largely unexplored.
The free electromagnetic field Hamiltonian is a sum of an infinite number of oscillator Hamiltonians, one for each mode. 
A first approach might be to apply the $n=1$ Roy-Sigh procedure to each mode separately.

\section*{\large{6. The n+1 marginal theorem: phase space Bell inequalities }}.
Quantum mechanics predicts only one of the $2^n$ probability densities $|\psi (\omega_1,..,\omega_n) |^2 $ in each measurement context,where
each $\omega_i$ can be $q_i$ or $p_i$.  
The 'n+1 marginal theorem'\cite{Auberson2002} asserts that at most $n+1$ of them can be simultaneously realized as marginals of a 
single positive definite phase space density.It is crucial in establishing that the Roy-Singh causal mechanics is 'maximally realistic'.
We illustrate the proof for $n=2$ case. 
Consider a four dimensional phase space with position variables $q_1 , q_ 2$ and momen-
tum variables $p_1 , p_2$. Let $\{\sigma _{ qq} (q_1 , q_ 2 ),\sigma _{ qp} (q_1 , p_ 2 ) ,\sigma _{ pq} (p_1 , q_ 2 ) ),\sigma _{ pp} (p_1 , p_ 2 ) \}$ be 
normalized probability distributions for an arbitrary pure quantum state $|\psi \rangle$,
\bea
\sigma_{qq} (q_1 , q_ 2 )= |\langle q_1 , q_ 2 |\psi\rangle |^2\>,\sigma_{qp}(  q_1 , p_ 2)= |\langle q_1 , p_ 2 |\psi\rangle |^2,\nonumber\\
\sigma_{pq}( p_1 , q_ 2 )= |\langle p_1 , q_ 2 |\psi\rangle |^2\>,\sigma_{pp}(p_1 , p_ 2  )= |\langle p_1 , p_ 2 |\psi\rangle |^2 ,
\eea
 or of the analogous form obtained by replacing $|\langle\xi|\psi\rangle|^2$ by $\langle\xi |\hat \rho |\xi \rangle $ for 
 a state with density operator $\hat \rho $.Is it possible to find a non-negative and normalized phase space density $\rho(\vec q,\vec p)$
of which $\sigma_{qq} , \sigma_{qp},\sigma_{pq} , \sigma_{pp}$ are marginals? i.e.
\bea
\int dp_ 1 dp_ 2 \rho(\vec q,\vec p) = \sigma_{qq} (q_1 , q_ 2 ) ,\nonumber\\
\int dp_ 1 dq_ 2 \rho(\vec q,\vec p) = \sigma_{qp} (q_1 , p_ 2 ) ,\nonumber\\
 \int dq_ 1 dp_ 2 \rho(\vec q,\vec p) = \sigma_{pq} (p_1 , q_ 2 ) , \nonumber\\
 \int dq_ 1 dq_ 2 \rho(\vec q,\vec p)= \sigma_{pp} (p_1 , p_ 2 ).\label{marginals}
 \eea

In $2n$ dimensional phase space, the Wigner function
\begin{eqnarray}
 W(\vec q,\vec p,t)= \frac{1} {(2 \pi)^n} \int d\vec y \exp (i \vec p. \vec y) \nonumber \\
  \langle \vec q-\vec  y/2 |\psi (t) \rangle \langle \psi(t) |\vec q+ \vec y/2 \rangle 
\end{eqnarray}
satisfies all these marginal conditions.
However, Hudson proved the theorem \cite{Hudson} that 
the only pure states with non-negative Wigner functions are those with Gaussian wave functions in $\vec q$-space.
Hence,in general, the Wigner function is not a solution to the problem we posed.

Phase space Bell inequalities come to the rescue \cite{Auberson2002}.
Consider the functions $r(q_ 1 , q_ 2 ), s(q_ 1 , p_ 2 ), t(p _1 , q_ 2 ) , u(p_ 1 , p_ 2 )$, defined by
\bea
r(q _1 , q_ 2 ) = sgn F_ 1 (q_ 1 ) sgn F_ 2 (q_ 2 ) ,\nonumber\\
s(q_ 1 , p_ 2 ) = sgn F_ 1 (q_ 1 ) sgn G_ 2 (p_ 2 ) ,\nonumber\\
t(p_ 1 , q_ 2 ) = sgn G_ 1 (p_ 1 ) sgn F_ 2 (q_ 2 ) ,\nonumber\\
u(p_ 1 , p_ 2 ) = -sgn G_ 1 (p_ 1 ) sgn G_ 2 (p_ 2 ) ,
\eea
where $F_ 1 , F_ 2 , G_ 1, G_ 2$ are arbitrary nonvanishing functions. Then, it is obvious
that for every $( q_ 1 , q_ 2 , p_ 1 , p_ 2 )$.
\be
r(q_ 1 , q_ 2 ) + s(q_ 1 , p_ 2 ) + t(p_ 1 , q_ 2 ) + u(p_ 1 , p_ 2 ) =\pm 2 .\label{rstu}
\ee
Suppose that a non-negative normalized phase space density $\rho(\vec q,\vec p)$ satisfying the four marginal
conditions (\ref{marginals} )exists. Multiplying eq.(\ref{rstu}) by  $\rho(\vec q,\vec p)$ and integrating over phase space,
we deduce the phase space Bell inequalities,
\be
|S|\leq 2\>, \label{BellIneq}
\ee
where,
\bea
S\equiv \int dq_ 1 dq_ 2 r(q _1 , q_ 2 )\sigma_{qq} (q_1 , q_ 2 ) \nonumber\\
+\int dq_ 1 dp_ 2 s(q_ 1 , p_ 2 ) \sigma_{qp} (q_1 , p_ 2 )\nonumber\\
+\int dp_ 1 dq_ 2 t(p_ 1 , q_ 2 ) \sigma_{pq} (p_1 , q_ 2 ) \nonumber\\
+\int dp_ 1 dp_ 2 u(p_ 1 , p_ 2 ) \sigma_{pp} (p_1 , p_ 2 ) .
\eea
 A long calculation establishes that the inequality (\ref{BellIneq}) is maximally violated  for
\bea
&|\psi\rangle_\pm (q_ 1 , q_ 2 ) = [1 \pm e ^{i\pi/4} sgn(q_ 1 )sgn(q_ 2 )]/ [2\sqrt 2]  \nonumber\\
&\times  h_ L (|q_ 1 |)h_ L (|q _2 |);h_ L (q)\equiv \theta (L -q)/\sqrt{(q + 1) \ln(L + 1)} ,\nonumber\\
&F_ i (q_ i ) = q_ i ,   G_ i (p_ i ) = p_ i ,   (i = 1, 2)  \nonumber
\eea
which yields $S \rightarrow \pm 2\sqrt 2$, for $L \rightarrow \infty $. This proves the
$n + 1$ marginal theorem for $n = 2$. Generalizations of the proof to arbitrary $n$ are also given in \cite{Auberson2002}.
 More general optimal wave functions for quantum optical tests have been constructed by  Wenger et al \cite{Grangier}.
 
\section*{\large{7.Bell inequality violations by states with positive Wigner function}}
  Banaszek et al \cite{Banaszek} made the brilliant discovery, missed by Bell, that EPR-Bell non-locality 
  depends not only on the quantum state but also on the observables measured.This work adds an essential 
  dimension to the foundations of quantum theory.
In his 1964 paper on the EPR paradox, Bell never used the original EPR state ,Eq.(\ref{EPR state}),
 perhaps because of its non-normalizability. Curiously, he returned to it in 1986 \cite{Bell1986}. 
 He wrote that the EPR state has no non-locality problem , because it admits a positive definite  Wigner function \cite{Wigner1932},
 which provides a local classical model of joint probability of position and momentum .  
 The two mode squeezed vacuum (TMSV) state produced by non-degenerate optical parametric amplification (NOPA) is a regularized EPR state:
 \begin{equation}
  |TMSV>= exp(\zeta^*a_1a_2-\zeta a_1^\dagger a_2^\dagger)|0>,\>\zeta =r\>exp (i\theta) .\nonumber
  \end{equation}
  For $\theta =\pi$, defining quadrature operators ,
  \be
  \hat {q}_k=\frac{ a_k+a_k^\dagger} {\sqrt{2} };\> \hat {p}_k=\frac{ a_k-a_k^\dagger} {i \sqrt{2} }
  \ee
  
  \begin{eqnarray}
 && <q_1,q_2|TMSV>=\pi ^{-1/2} exp\big(-exp(-2r) \frac{ (q_1+q_2)^2} { 4}\nonumber\\
 &&-exp(2r) \frac{ (q_1-q_2)^2} { 4}\big)\rightarrow _{r\rightarrow \infty } 2 exp(-r)\delta (q_1-q_2) \>.\nonumber
  \end{eqnarray}
Being Gaussian, it has a non-negative Wigner function , and confirms Bell's claim, provided only  $q,p$ measurements
are considered. However, Banaszek et al \cite{Banaszek} 
 showed that this EPR state, inspite of a positive Wigner function, violates Bell inequalities on correlations of 
 phase space displaced parity operators:
 \bea
 A_i(\alpha_i)=D_i(\alpha_i)P_i D_i (\alpha_i^*)\>, P_i=\int_{-\infty}^{\infty} dq_i |q_i\rangle \langle - q_i| \nonumber\\
 D_i(\alpha_i)=exp(\alpha_i a_i^\dagger -\alpha_i^* a_i),\alpha_i=(q_i+ip_i)/\sqrt 2\>.
 \eea
 Being unitary transforms of the parity operator,the  $ A_i(\alpha_i)$ have eigenvalues $\pm 1$ just like the spin components used by Bell,
 and the standard Bell-CHSH inequalities with $ a,b,a',b'\rightarrow \alpha,\beta,\alpha',\beta'$ follow for the 
 correlations $P(\alpha,\beta)=\langle A_1(\alpha) A_2(\beta)\rangle $. Using the representation of the Wigner function 
 for $n-$dimensional configuration space \cite{Royer}
 in terms of the $A_i$, for the TMSV state with $\theta =\pi$,
 \bea
 W(\vec q,\vec p)= (\pi \hbar)^{-2}Tr \hat \rho A_1 (\alpha_1)  A_2 (\alpha_2) \nonumber\\
   = (\pi \hbar)^{-2}exp\{-cosh (2r)(q_1^2+q_2^2+p_1^2+p_2^2)\nonumber\\
   +2 sinh (2r)(q_1q_2-p_1p_2) \},
\eea
Banaszek et al \cite{Banaszek} could reach quantum values upto 2.19 for the left-hand side of Eqn. (\ref{Bell-CHSH}) for special choices of 
phase space displacements,$r\rightarrow \infty$ in violation of local reality. Chen et al \cite{Chen} confirmed their conclusion by reaching a value $2\sqrt 2$ for the left-hand side of Eqn.(\ref{Bell-CHSH}) 
for the same state by using $P(a,b)=\langle \vec \sigma _1.\vec a \vec \sigma _2.\vec b  \rangle $, where the $\vec \sigma_i $ are 
pseudo-spin operators related to the parity operator. However ,for any canonical pair ,the Wigner function ,when positive definite,could 
provide a possible joint probability distribution of operators such as $|q\rangle \langle q|,|p\rangle \langle p|$ which are diagonal in position 
or momentum. The parity operator fails this criterion.

I thank Vandana Nanal, Editor, Physics News for the invitation to write this article and for suggestions on improving the manuscript,
I also thank G. Auberson, G. Mahoux and Virendra Singh for valuable suggestions on the manuscript.

\end{document}